\documentclass[12pt]{article}
\begin{document} 
\baselineskip 18pt
\newcommand{\Tr}{\mbox{Tr\,}}
\newcommand{\beq}{\begin{equation}}
\newcommand{\eeq}[1]{\label{#1}\end{equation}}
\newcommand{\bea}{\begin{eqnarray}}
\newcommand{\eea}[1]{\label{#1}\end{eqnarray}}
\renewcommand{\Re}{\mbox{Re}\,}
\renewcommand{\Im}{\mbox{Im}\,}
\begin{titlepage}
\hfill  CERN-TH/99-46 BICOCCA-FT-99-05 IMPERIAL/TP/98-99/42
NYU-TH/98/2/03\vskip .01in \hfill hep-th/9903026 
\begin{center}
\hfill
\vskip .4in
{\large\bf Confinement and Condensates Without Fine Tuning in Supergravity
Duals of Gauge
Theories}
\end{center}
\vskip .4in
\begin{center}
{\large L. Girardello$^{a,b}$, M. Petrini$^c$, M. Porrati$^{a,d}$ and
A. Zaffaroni$^a$\footnotemark}
\footnotetext{e-mail: girardello@milano.infn.it, \,m.petrini@ic.ac.uk,\,
massimo.porrati@nyu.edu,\, alberto.zaffaro-\\ ni@cern.ch}
\vskip .1in
(a){\em Theory Division CERN, Ch 1211 Geneva 23, Switzerland}
\vskip .1in
(b){\em Dipartimento di Fisica, Universit\`a di Milano-Bicocca and INFN, 
Sezione di Milano, Italy\footnotemark}
\footnotetext{Permanent Address}
\vskip .1in
(c){\em Theoretical Physics Group, Blackett Laboratory,
Imperial College, London SW7 2BZ, U.K.}
\vskip .1in
(d){\em Department of Physics, NYU, 4 Washington Pl.,
New York, NY 10003, USA\footnotemark}
\footnotetext{Permanent Address}
\end{center}
\vskip .4in
\begin{center} {\bf ABSTRACT} \end{center}
\begin{quotation}
\noindent
We discuss a solution of the equations of motion of five-dimensional
gauged type IIB supergravity that describes confining $SU(N)$ gauge theories at
large $N$ and large 't Hooft parameter. We prove confinement by computing the
Wilson loop, and we show that our solution is generic, independent of most of
the details of the theory. In particular, the Einstein-frame
metric near its singularity, and the condensates of scalar, composite
operators are universal. Also universal is the discreteness of the glueball
mass spectrum and the existence of a mass gap.
The metric is also identical to a generically confining solution recently
found in type 0B theory. 
\end{quotation}
\vfill
CERN-TH/99-46 \\
February 1999
\end{titlepage}
\eject
\noindent
\section{Introduction}
The duality between supergravity theories and strongly coupled field theories
at large $N$ and large 't Hooft parameter~\cite{'t} applies both to
conformal~\cite{malda,gkp,w1} and non-conformal field theories~\cite{imsy,w2}.

In order to be able to trust the supergravity approximation, one finds,
quite generally, that the 't Hooft coupling, $x=Ng_{YM}^2$, must be large. In
non-conformally invariant theories, where $x$ runs, this is to be understood
as $x$ at the ultraviolet cutoff. This means that the dynamical regime of
gauge theories that can be described via supergravity duals is not the
usual one, where $x\ll 1$ at the UV cutoff.
Nevertheless, it is believed that strongly-coupled
gauge theories share some qualitative properties with their weakly coupled
relatives; in particular, confinement and mass gap~\cite{w2}.
This belief has given rise to a blossoming of works where
(super)gravity solutions are found, that describe higher-dimensional gauge
theories compactified to 4-d~\cite{r}, deformations of
N=4 4-d supersymmetric gauge theories~\cite{ks}, and deformations of
non-supersymmetric conformal theories~\cite{kt,m}.
In latter example, the dual of the strongly-coupled gauge theory is a type 0,
non-supersymmetric string theory.

A question posed by the existence of so many different ways of obtaining
theories that are expected to have the same infrared behaviour is whether
there exists a framework that allows to describe all the confining solutions
at once. A slightly less ambitious objective is to find a framework
that describes generically confining deformations of N=4, 4-d super
Yang-Mills.

In this paper, we describe such framework: 5-d gauged supergravity.
We find a generic, universal solution of the equations of motion of gauged
supergravity, that 
does not depend on the detailed form of its action. Our results
exceed our expectations. Indeed, we find that, in our solution,
the generic IR behaviour of the metric  
is universal, and agrees not only with all 
known confining solutions of type IIB
supergravity, corresponding to softly broken N=4 super Yang-Mills, but
also with solutions of type 0 supergravity. 

Our scenario has some similarities with the one proposed in \cite{poly,poly2} 
for describing 
the universality class of confining strings, using five-dimensional
non critical string theories. As a matter of fact our solution agrees in the 
IR with the 
one found in \cite{poly} for the simplest background with only the metric and 
the dilaton. It would be interesting to check if more complicated confining
solutions, obtained by non-critical string theories, have the same 
 IR behaviour as the one found in this paper.

We discuss the main features of our solution, as discreteness of the
glueball mass spectrum and the existence of a mass gap. We present the
computation of the string tensions and of the Wilson loop, and, in a 
concrete example, we show explicitly the
area law for the quark-antiquark Wilson loop.  
We also find a general formula for condensates of scalar composite
operators. This formula depends only on the near-singularity form of the
metric.

This paper is organised as follows. In Section 2, we describe the general
model-independent results of our analysis. This Section describes
the computation of the near-singularity metric, Wilson loop, glueball
mass spectrum, and 
condensates. Section 3 applies the formalism of Section 2 to a specific
example in 5-d gauged supergravity. Section 4 contains our conclusions.
Technical aspects of the computations in Sections 2 and 3 are confined to two
Appendices.
\section{General Results}
The four-dimensional N=4 $SU(N)$ super Yang-Mills theory is a
superconformal field theory whose supergravity dual is type IIB supergravity
compactified on $AdS_5\times S_5$~\cite{malda}. Type IIB supergravity
contains all deformations of N=4, with
dimensions that remain finite in the large-$x$ limit; they were found
in~\cite{krv}. Some of these deformations also exist in type IIB
{\em dimensionally reduced} to five dimensions on $AdS_5\times S_5$, that is,
in 5-d gauged N=8 supergravity~\cite{grw1,grw2}.

Gauged N=8 supergravity has gauge group $SU(4)$ and 42 scalars. By the
AdS/CFT correspondence, these scalars correspond to deformations of the N=4
super Yang-Mills theory with dimension 2, 3, and 4. The scalar potential
of 5-d gauged supergravity has several stationary points. All stationary
points that satisfy the Breitenlohner-Freedman stability bound~\cite{bf}
correspond to local, unitary, four-dimensional conformal field theories.
Some of these stationary points have been studied in~\cite{gppz,dz,kpw}.

It is widely believed that each solution of the equations of motion of
5-d gauged supergravity can be promoted to a solution of 10-d type IIB
supergravity. Arguments in favour of this statement can be found
in~\cite{gppz,kpw,df}; moreover, this is a theorem in the largely analogous
case of 4-d gauged $SO(8)$ supergravity~\cite{dn}.

In a previous paper~\cite{gppz}, we went beyond the study of conformal fixed
points, and we found solutions of 5-d gauged supergravity that interpolate
between two stationary points of the scalar potential. The 5-d metric of such
solutions has two asymptotically-$AdS_5$ regions, describing the UV and IR
conformal fixed points of the gauge theory.

In this paper, instead, we want to find solutions of 5-d gauged supergravity
with only one asymptotically-$AdS_5$ region, corresponding to the UV N=4 super
Yang-Mills $SU(N)$ theory. We will find that the IR behaviour of this metric
is universal.

We are looking for a metric that preserves 4-d Poincar\'e invariance.
This restricts its form to
\beq
ds^2= dy^2 + e^{2\phi(y)}dx^\mu d x_\mu, \;\;\;\mu=0,1,2,3.
\eeq{m1}
We always work in the 5-d Einstein frame; $AdS_5$ with radius $R$ corresponds
to $\phi= y/R + {\rm const}$. As in~\cite{gppz}, we associate larger
energies with increasing $y$. This means that we
must look for solutions that approach $AdS_5$ when $y\rightarrow \infty$.

We also allow for a $y$-dependence in some of the 42 scalars we have in our
theory. The boundary conditions for the scalars is that they approach
the $SO(6)$, N=8 invariant stationary point. This means that for
$y\rightarrow \infty$ the dilaton is a constant, while all the other scalars
vanish.

At large $y$, the behaviour of the scalars and the Einstein metric is
complicated and non-universal, {\em but their infrared behaviour
is universal}, for a wide class of solutions.

The 5-dimensional supergravity action for scalar fields coupled
to the metric~(\ref{m1}) reduces to
\beq
L =\sqrt{-g}\left[
{R\over 4} + {1\over 2}\sum_a (\partial_y \lambda^a)^2 +V(\lambda^a)\right].
\eeq{mn9}
Here we have assumed canonical kinetic term for all scalars. This can always
be achieved by an appropriate field redefinition. As we shall see
shortly in our concrete example, the kinetic term is already canonical with a standard
choice of variables. Einstein's equations and the equations of motion of
the scalars following from (\ref{mn9}) are:
\beq
\partial_y^2\lambda^a + 4\partial_y\phi \  \partial_y\lambda^a= {\partial V \over \partial \lambda^a},
\;\;\; 6(\partial_y\phi)^2=\sum_a (\partial_y\lambda^a)^2 -2V.
\eeq{m2}

Eqs.~(\ref{m2}) have a universal runaway solution, independent of the detailed
form of the potential. It is quite similar to the ``singular instanton''
solution found by Hawking and Turok~\cite{ht}. Namely, let us assume that
near $y=a$ the scalar fields and the metric diverge such that the potential
term in Eqs.~(\ref{m2}) is irrelevant. Then, one find a universal behaviour for
both $\phi$ and $\lambda^a$ near $y=a$:
\beq
\lambda^a= -  K^a \log |y-a| +{\rm constant}, \;\;\; \phi= {1\over 4} \log
|y-a|
+{\rm constant}, \,\,\, \sum K^{a\, 2} = {3\over 8}.
\eeq{m3}

The metric, in particular, has a solution-independent behaviour:
\beq
ds^2= dy^2 + |y-a|^{1/2}dx^\mu dx_\mu.
\eeq{m4}
For notational simplicity, we will put $a=0$ in the following. 

The assumption
 that the potential is irrelevant is justified whenever $V$ is
polynomial in $\lambda^a$. In gauged supergravity, typically, when scalars
are canonically normalised, the potential is exponential in some of them.
This means that the constants $K^a$ in Eq.~(\ref{m2}) must obey some
model-dependent inequalities.

Eq.~(\ref{m4}) agrees with
the behaviour of the 5-d Einstein-frame metric found in
refs.~\cite{ks,m}. Agreement with ref.~\cite{ks} is encouraging, since
that paper studies a deformation of type IIB theory on $AdS_5\times S_5$;
i.e. a model that falls into the class we are studying.
More surprising is agreement with ref.~\cite{m}. After all, that paper studies
a compactification of the non-supersymmetric type 0 superstring. 
Moreover, the form of the metric
also agrees with that found in \cite{poly}, using general arguments based on 
five-dimensional non-critical string theories, for a background with only the
metric and the dilaton\footnote{The relevant behaviour  for this solution 
is determined by eq. (26) of \cite{poly}. Notice that we want
to interpret $\phi =0$ not as the boundary, as in \cite{poly}, 
but as the IR, as in the AdS-revised philosophy discussed in \cite{poly2}.}. 
The common behaviour of the
IR metric in all these different cases, type IIB, type 0 and non-critical strings, has a technical reason: all these 
models can be reduced 
to some effective theory for five-dimensional scalars. The fact
that in all these theories one finds the same singularity 
in the metric may mean that
the result we obtained is truly universal, and largely insensitive to the
UV theory used to derive the IR confining behaviour. 

We now discuss what are the characteristics  of solutions that describe 
confining theories. The  asymptotic behaviour in Eq.~(\ref{m4}) 
does not   completely specified the model. We must give the
expressions for the tensions of the fundamental string and of the Dirichlet
string.
Since we made several Weyl rescaling with the purpose of reducing the
ten-dimensional equations of motion in the string frame to the five-dimensional
Einstein frame, we  expect that the five-dimensional tensions depend on the
supergravity scalars. These tensions are, in general, model-dependent. We
compute below
the tensions for a particular solution. However, from the computational details
for the specific solution and the form of the five-dimensional supergravity, we
expect a general power behaviour for the tensions. If we single out the
particular scalar $\lambda^0$
as the five-dimensional dilaton, we have,
\beq
T_{F1}=y^{-\sum d_a K^a -  K^0},\;\;\;
T_{D1}=y^{- \sum d_a K^a +  K^0},
\eeq{m5}
where the coefficients $d_a$ depend on the specific model. This behaviour
is in  agreement with that found in ref.~\cite{ks,m}. 
Equipped with this result, we can 
study whether the theory confines by computing a Wilson loop, according to
the original proposal in \cite{maldaloop}. The Wilson loop associated to a 
given boundary contour $C$ is obtained by 
minimising the classical action for a string free to move in the 
five-dimensional space but with endpoints on $C$. 
As reviewed in Appendix A, the knowledge of the asymptotic behaviour of 
the metric and the tensions is sufficient for extracting qualitative 
information about the Wilson loop. The result is that,
in the range of the allowed values for $K^a$, it is possible to
find different behaviours for the quark-antiquark potential, namely
confinement as well
as screening. However, confinement is generic  for this class of solutions, 
in the sense that there is a large range of values of $K^a$ for which
quarks confine and monopoles are screened. This is similar to what found
in \cite{m}. We exhibit below the detailed
calculation in a specific example. Moreover, as we will see,
even if the numerical details about the range of allowed values
for $K^a$ are model-dependent, the physical picture is always the same.
The asymptotic behaviour of the metric~(\ref{m4}) conspires with the form
 of the tensions~(\ref{m5}) to create a barrier for the fundamental string,
which therefore cannot enter in the deep IR region. Since the string is 
forced to stay near to the boundary, an area law behaviour for the
Wilson loop is expected \cite{w2}.

Some comments about the validity of the solution are in order. Since the
metric and the kinetic terms for scalars blow up in the IR, we expect that the
supergravity solution is significantly corrected
by the higher derivatives terms in the Lagrangian. 
This problem is common to all the proposed supergravity solutions 
for confining theories \cite{ks,m}.
We may expect that corrections are mildened by the fact that the Weyl tensor
vanishes for our background \cite{kt}.
Moreover, it was shown in
\cite{m}, which contains solutions closely related to ours, 
that there exist particular values of the parameters for which the
corrections are suppressed. In the best of possible worlds, the
corrections will simply slightly change the range of parameters, and we saw
that the physics of our solution is not sensitive to such details. Moreover, 
confinement is obtained when the fundamental string does not probe the
far IR region, whose details become therefore irrelevant. It is possible then 
that the particles wave-functions and the fundamental string world-sheet,
needed for computing the mass spectrum and the Wilson loops,
always live in a region where curvature and higher
 derivatives terms are still under control.

We now discuss the existence of a mass gap and the glueball spectrum.
This is done as in~\cite{w2} by looking for normalizable solutions of the
equation of motion of a minimally coupled massless scalar
in the metric~(\ref{m1})
The action of the scalar is
\beq
S=
\int_0^\infty dy \left(e^{4\phi}|\partial_y \psi|^2 + k^2 e^{2\phi}|\psi|^2
\right), \;\;\; k^2=k^\mu k_\mu.
\eeq{mm1}
Integrating by part Eq.~(\ref{mm1}), and using the equation of motion and
$e^{4\phi}=0$ at $y=0$ we find $S=0$ for the normalizable solution, which
approaches a constant at $y=0$. 
Therefore, $k^2$ must be strictly negative, proving the existence of a 
mass gap. Discreteness of the spectrum follows from requiring normalizability
at large $y$~\cite{w2}.

The equation of motion derived from action~(\ref{mm1}) is
\beq
-\partial_y^2\psi -4\partial_y\phi \ \partial_y \psi +k^2e^{-2\phi}\psi=0.
\eeq{mm2}
The spectrum of glueballs can be studied using a WKB 
approximation~\cite{mina}.
Standard calculations lead to the following form for the WKB wave function:
\bea
\psi(y)&=& C [-2\phi(y) \pm i p(y)]^{-1/2}
e^{-2\phi(y)\pm i\int_a^y dy p(y)},\;\;\;
0<y<b\\
\psi(y)&=& C' [2\phi(y)+ |p(y)|]^{-1/2}
e^{-2\phi(y)-\int_y^a dy |p(y)|}, \;\;\; y<a,\\
\psi(y)&=& C'' [2\phi(y)+ |p(y)|]^{-1/2}
e^{-2\phi(y)-\int_b^y dy |p(y)|}, \;\;\; y>b. 
\eea{mm3}
Here, $p(y)=\sqrt{-k^2e^{-2\phi(y)} -4\phi^2_y(y)}$, $C,C',C''$ are
constants, and $a,b$ are the classical turning points, $p(a)=p(b)=0$.
The existence of these points follow from the form of $p(y)$ at small and 
large $y$. At small $y$, it is:
\beq
p(y)= -k^2 y^{-1/2} -{1\over 4} y^{-2}, \;\;\; y \ll 1.
\eeq{mm4}
Notice that this expression is universal, since it depends only on the 
near-singularity metric in Eq.~(\ref{m4}). At large $y$, $p(y)$ is also 
universal, since there the metric approaches the AdS form: $\phi(y)=y$;
thus,
\beq
p(y)= -k^2 e^{-2y} - 4.
\eeq{mm5}
At $y=0$, $p(y)$ diverges to $-\infty$; at intermediate $y$, $p(y)$ is positive
when $k^2<0$; at large $y$, $p(y)$ becomes negative again. This proves the
existence of turning points.
As usual in the case of a bound classical trajectory, 
the bound state spectrum 
is discrete. In the WKB approximation, it is given by the Bohr-Sommerfeld
quantisation rule~\cite{ll}
\beq
\int_a^b dy\sqrt{E_n 
e^{-2\phi(y)} -4\phi_y^2(y)}=\pi (n +\alpha),\;\;\; E_n=-k^2,
\eeq{mm6}
with $\alpha$ a constant of order one.
Eq.~(\ref{mm6}) is the starting point for numerical
and analytical studies of the glueball spectrum.
In this paper, though, we do not want to join this thriving cottage
industry; instead, we want to exhibit general, model independent, features
of supergravity duals of gauge theories.
Thus, we shall abandon the study of the glueball spectrum, and discuss another 
model-independent feature of our solution, namely, the existence of 
condensates.
 
The condensate of a field theoretical
operator $O_a$ that couples to the source $\lambda^a$ reads~\cite{gkp,w1}:
\beq
\langle O_a\rangle ={\delta S_{5d} \over \delta \lambda^a}.
\eeq{mm7}
Here $S_{5d}$ is the action of 5-d gauged supergravity, computed on the
solution of the equations of motion~(\ref{m2}). Thanks to
the equations of motion, the right hand side of Eq.~(\ref{mm7}) is a total
derivative:
\beq
\langle O_a\rangle = \left[
{1\over 2}e^{4\phi(y)}M_{ab}(\lambda)\lambda^b_y\right]_0^\infty.
\eeq{mm8}
Unlike in Eq.~(\ref{m2}), here we allow for a non-canonical kinetic
term. Whenever the potential $V$ can be neglected, the universal runaway
solution in Eq.~(\ref{m3}) still holds. Near $y=0$ it reads
\beq
e^{4\phi(y)} M_{ab}(\lambda)\lambda^b_y={\rm constant}\equiv -K_a.
\eeq{mm9}
In Eq.~(\ref{mm8}) the upper boundary diverges at $y=\infty$ and must be
subtracted. This UV regularization is determined by requiring that
condensates vanish for geometries with two asymptotic AdS regions. In that
case, the coordinate $y$ runs from $y=-\infty$ to $y=+\infty$, and the
contribution of the lower boundary to Eq.~(\ref{mm8}) vanishes, while
the contribution of the upper boundary is still divergent.
After UV regularization, Eq.~(\ref{mm8}) reduces to a very simple form:
\beq
\langle O_a\rangle={1\over 2}K_a.
\eeq{mm11}
Notice that this formula for the condensate is independent of all the fine
details of the metric, in particular, its behaviour at large $y$. Moreover,
it is finite without any {\em ad-hoc} IR regularization, and also gives a
physical meaning to the constants of integration $K_a$.
\section{An Example}
Let us apply now the machinery developed in the previous Section to a
specific example.
A simplified theory with only two scalars it is already enough to describe  the
general behaviour of confining theories, and allows us to make contact with the
solutions found in \cite{ks,m}.

It is easy to make contact with the solution found in \cite{ks} . We are
studying the same problem in the dimensionally reduced theory. The solution in
\cite{ks} corresponds to a single running field, the five-dimensional dilaton,
which has zero potential \cite{grw2}. We therefore expect an IR metric as in
Eq.~(\ref{m4}). A simple Weyl rescaling indeed shows that the solution in
\cite{ks} has the IR behaviour given by Eq.~(\ref{m4}).
The tensions also agree with the discussion in the previous Section.

More general ten-dimensional solutions, corresponding to perturbations of
$AdS_5\times S^5$ with  some other fields, and breaking the $SU(4)$ invariance,
have not yet been found. These solutions are quite natural to consider, since
most of them correspond to perturbations of super
N=4 Yang-Mills with mass terms
for scalars and fermions. The technical difficulty of finding a full
ten-dimensional solution can be easily overcome if we are interested in the
less complete but still quite interesting corresponding five-dimensional
solution. In ten dimensions, $SU(4)$ non-invariant fields are automatically
sources for the dilaton, which then starts to run.
We can mimic this solution in
five dimensions by considering the above equations of motions for two scalars,
the dilaton (which does not appear in the potential) and a scalar with a
non-trivial potential. We consider, for simplicity, the lucky case in which all
other scalars can be consistently set to zero. The
solution in Eq.~(\ref{m3}) depends on two constants, $K$ and $K_0$, with the
constraint $K^2+K^2_0=3/8$. A potential for $\lambda$, which in supergravity
generally is of the asymptotic form $e^{c\lambda}$, is irrelevant if and only
if
\beq
0 < cK < 2, \,\,\, K^2 +K_0^2 ={3\over 8}.
\eeq{m7}
Generically,
this equation can be satisfied for certain values of $K$, in which case the IR
metric assumes the form given in Eq.~(\ref{m4}). We could also try to consider
a solution with a fixed value of the dilaton. In this case, it is not obvious
that the potential is irrelevant; it must happen that $c\sqrt{3\over 8}< 2$
and, as we will see, this is not the case, in general.

The solutions found in \cite{m} do not obviously fall in the class of examples
we are considering; indeed, they are not
obtained from an N=4 UV fixed point. The source for the dilaton is, instead, a
non-zero tachyon  field, which exists only in type 0 theories.
However, the type
0 equations of motions also reduce to a certain Lagrangian for a set
of scalars with an effective potential. Under these circumstances, it is
not surprising that the solution found in \cite{m} has the same generic IR
behaviour as our solution.

To study a concrete example, let us consider a deformation of the N=4
Yang-Mills theory that corresponds, in N=1 notations,  to a mass term for the
three chiral multiplets $X_i$. The theory
 flows in the IR to pure N=1 Yang-Mills, which confines. To obtain
the standard N=1 pure Yang-Mills with fixed scale $\Lambda$ we need a fine
tuning of the UV parameters, in which the mass $m$ diverges while the t'Hooft
coupling constant, $x$, goes to zero as a logarithm of $m$. This is outside
the regime of validity of supergravity, which requires a large $x$, but we
may still expect to see from supergravity the qualitative properties of the
theory, as, for instance, confinement.
The supergravity mode corresponding to a mass term
for the chiral multiplets was identified in \cite{flz}. The $42$ scalars of
the five-dimensional supergravity transform as
$\underline{1},\underline{20},\underline{10}$ under $SU(4)$. The explicit form
of the CFT operators to which these supergravity fields couple can be found,
for example, in \cite{gppz}. 
The mass term for the chiral multiplets appears in the
decomposition $\underline{10}\rightarrow \underline{1}+\underline{6}
+\underline{3}$ of $SU(4)$ under $SU(3)\times U(1)$. A non-zero vev for the
$SU(3)$ singlet contained
in this decomposition was studied in \cite{gppz,dz}. It corresponds to
a deformation of N=4 super Yang-Mills that
leads to a non-supersymmetric, conformal
IR fixed point. A non-zero vev for $\underline{6}$, when represented as a
complex symmetric matrix, $m_{ij}$, corresponds to the N=1 supersymmetric mass
deformation $\int d^2\theta m_{ij} X_i X_j + {\rm c.c.}$. Since we want
to compute and to control the asymptotic
behaviour of the potential, we need to reduce the number of scalars which we
turn on. If we are not careful, a non-zero vev for the scalar $m$ will
induce non-zero vevs for other scalars as well, due to the existence of linear
couplings of $m$ to the other fields in the potential. This is indeed the case for
generic $m_{ij}$. However, if we further impose $SO(3)$ symmetry,
by taking an $m_{ij}$ proportional to the identity matrix, a
simple group theory exercise shows that all the remaining fields can be
consistently set to zero.

The five-dimensional action of the scalars, given in ref.~\cite{grw1},
is written in terms of a
$27\times 27$ matrix $U$, transforming in the fundamental representation of
$E_6$ and parametrising the coset $E_6/USp(8)$. $U$ can be written as $U=e^X,
X=\sum_a \lambda_a T_a$, where, in a unitary gauge, $T_a$ are the
generators of $E_6$ that do not belong to $USp(8)$. This matrix has exactly 42
real independent parameters, which are the scalars of the supergravity theory.
The Lagrangian has the form
\beq
L = \sqrt{-g}\left[-{R\over 4} - {1\over 24}\Tr (U^{-1}\partial U)^2 + V(U)
\right].
\eeq{m8}
 Since the rules for computing the potential were described in \cite{grw2},
and recently reviewed and applied in \cite{dz,kpw}, here we will recall only
the main points of the computation.
First, one chooses a convenient parametrisation for the coset manifold
representative which makes the kinetic terms canonical. A possible
parametrisation,
suitable for studying the breaking $SU(4)\rightarrow SU(3)\times U(1)$,
 has been extensively described in \cite{dz}.
 Secondly, one performs some gamma-matrix algebra. Conventions,
details and
computational tricks can be found in ref.~\cite{grw2}. A summary
of the relevant
formulae can be found in Appendix B.
The result for the diagonal scalar $m\delta_{ij}$ is
\beq
L =\sqrt{-g}\left[-
{R\over 4} + {1\over 2}(\partial m)^2 - {3\over 8R^2}(\cosh^2(2m)+7)\right],
\eeq{m9}
where $R$ is the curvature of $AdS_5$. The  above potential passes two crucial
checks: when expanded for small $m$ it reproduces the cosmological constant
of the $SU(4)$ symmetric AdS-vacuum, and the mass for the scalar $m$
--which is,
according to \cite{krv}, $-3/R^2$, the right value to be the source of a
dimension-three operator in the CFT. In any case, the asymptotic behaviour of
the potential  can be figured out even without a detailed computation.
With only one scalar and the above form for $U=e^{mT}$, where $T$ is a
particular generator of $E_6$, suitably normalised, the kinetic term is
automatically canonical.
Moreover, the potential is quartic in $U$ making quite plausible the asymptotic
behaviour $\sim e^{4m}$.

If we turn on only the field $m$,
and implement the constraint in Eq.~(\ref{m7}),
we easily discover that the potential cannot be neglected, and, therefore, the
solution in eq.~(\ref{m2}) does not describe well the IR physics. This is due
to the fact that the power of $e^m$ in the potential is too large.
As we already
discussed, this problem can be easily circumvented by turning on other scalars.
The natural choice  is to turn on the dilaton. We will call $\rho$ the
five-dimensional dilaton. It has a canonical kinetic term ${1\over
2}(\partial\rho )^2$,
and it does not appear in the potential. It is not obvious whether the
five-dimensional
solution preserves N=1 supersymmetry, as one may naively think. The
solution with only the dilaton was argued to break all supersymmetries
in~\cite{ks}. It is possible that some particular solution with non-zero
fields $m$ and $\rho$ preserves N=1 supersymmetry; the  safe way to
check this is to look at the supersymmetry variations of the fermions. However,
here we are not really interested in an N=1 solution. Rather, we want to prove
that for a generic,
not necessarily supersymmetric solution, the IR behaviour is
universal, and that confinement is expected for a large range of values of the
parameters $K_a$, without requiring fine tuning. The non-zero
dilaton induces some other $SO(6)$ invariant deformations of the N=4 UV fixed
point. Beside the obvious $F_{\mu\nu}^2$
term, the only other $SO(6)$-invariant supergravity field that may appear is
$F^4$. This is however irrelevant at the UV fixed point and does not explain
the solution in \cite{ks}. It was argued in the second paper in
ref.~\cite{ks} that
the usually neglected non-chiral operators, like the diagonal mass term for the
six scalars of the N=4 theory, may be also induced as deformations of the N=4
UV fixed point by the running dilaton solution,
and that they may play a role in
the field theory interpretation, despite their infinite anomalous
dimension\footnote{With a deformation of the form $\int d^2\theta m X_i X_i +
{\rm c.c.}$, we find in the Lagrangian also a diagonal mass term for the six
scalars of the N=4 theory, quadratic in the parameter $m$, and, therefore,
not seen in a linearization in $m$. The CFT operator associated to the
supergravity scalar $m$ is obtained by linearizing around the $AdS_5\times S^5$
solution, and it will never see the quadratic term. The appearance of this
non-chiral operator at the second order may be an hint for the inclusion of
stringy state, or simply due to an operator mixing.}.
Whether or not we accept this interpretation,
the solution with non-zero dilaton
and non-zero scalar $m$ is reasonably associated to a deformation of N=4
super Yang-Mills,
that renders massive at least all the scalars and fermions in the
chiral multiplets. In a genuine N=1 theory, the gaugino will remain
massless to give pure N=1 Yang-Mills.
In a non supersymmetric theory it will probably
become massive. In any case, we reasonably expect a confining theory in the IR.

To prove confinement we need to compute the Wilson loop and show that it
exhibits an area law. We first need to compute the string tension in five
dimensions.
The  tension of the fundamental string (or the D1-string) can be read from the
coefficient of the kinetic term for the NS-NS (or R-R) antisymmetric  tensor in
the ten dimensional Lagrangian in the Einstein frame,
\beq
{1\over T_{F1}^2}H^2_{NS\mbox{-}NS} + {1\over T_{D1}^2}H^2_{R\mbox{-}R}.
\eeq{m11}
A simple Weyl rescaling shows that this property is valid also in the
five-dimensional theory in the Einstein frame. We need, therefore, the kinetic
terms of the antisymmetric tensors in the five-dimensional supergravity.
The five-dimensional Lagrangian has a global $SL(2;R)$ symmetry, which
helps in identifying the various types of strings. There are 12 antisymmetric
tensors $B_{\mu\nu}^{I\alpha}, I=1,...,6, \alpha =1,2$, transforming in the
$\underline{6}$ of $SU(4)$, and in the two-dimensional representation of
$SL(2;R)$.
We interpret the index $I$ as signalling that our five-dimensional string
remembers its ten-dimensional origin. The position of the string
on the five-sphere,
indeed, must be specified in a bona-fide
ten-dimensional computation. We will consider
the case in which the kinetic terms for $B^{I\alpha}$ are diagonal in $I$.
The second index, $\alpha$,
specifies a whole multiplet of strings transforming
under $SL(2;R)$, as it is expected from the S-duality of N=4 super Yang-Mills.
The antisymmetric-tensor kinetic term was written in \cite{grw2} in a first
order formalism:
\beq
\epsilon_{\alpha\beta}B_{I\alpha}\wedge dB_{I\beta} + A_{I\alpha,J\beta}
B_{I\alpha}\wedge *B_{J\beta}.
\eeq{m12}
$A_{I\alpha,J\beta}$ is a symmetric matrix which depends on the scalars and
whose explicit expression is given in Appendix B. 

The problem of writing down the BPS formula, which relates the tension of
the strings to their charges, is non trivial, especially in a gauged 
supergravity, where the fields $B_{I\alpha}$ are massive. In this
paper we are taking the quite plausible point of view of believing that the
five-dimensional
$SL(2;R)$ is the same as the ten-dimensional one and it is broken to
$SL(2;Z)$
by non-perturbative effects. We also trust naive dimensional reduction
arguments from ten dimensions\footnote{The general discussion in the
previous section would not be affected by the discovery of the na\"{i}vety
of this argument. And, for the particular example we are discussing, 
only few details would change; certainly not the general conclusions.}. For a
diagonal matrix $A$ we simply have
\beq
B_1\wedge dB_2 + a_1|B_1|^2 + a_2|B_2|^2.
\eeq{m13}
Now, if we
choose one of the fields (say $B_1$) and solve the equation of motion of
$B_2$, we can recover the standard action for a massive antisymmetric tensor,
which is expected in KK reduction \cite{krv}:
\beq
{1\over a_2}|dB_1|^2 + m |B_1|^2.
\eeq{m14}
When $A$ is not diagonal, one must be careful in going from the first order
formalism to the second order one. 
One can easily see that the square-root of the eigenvalues of the
matrix $A_{I\alpha,J\alpha}$, restricted to diagonal $SL(2,R)$ indices,  
gives the
tensions of the strings. In the parametrisation used in this paper,
the square of the  tension of the fundamental string is given by 
an eigenvalue of 
$A_{I1J1}$, while the square of the tension of the D1-string is given
by an 
eigenvalue of 
$A_{I2,J2}$. The matrix $A$ is
essentially the square of the matrix $U$; the precise formula is given in
Appendix B.
While the potential is independent of the dilaton, $A$ crucially depends on it.
It is easy to compute the eigenvalues of $A$ for the maximally supersymmetric
vacuum corresponding to N=4 super
Yang-Mills, when all the scalars but the dilaton
are zero. The detailed computation can be found in Appendix B. The not
surprising result is that the two eigenvalues of $A$, $(a_1, a_2)$, are exactly
$(e^{2\rho},e^{-2\rho})$. The tensions can be computed without difficulty
even when other scalars are non-zero. What one finds is that, due to
the quadratic dependence on $U$, the contribution of $m$ to both tensions
has the asymptotic form $\sim e^{m}$. 
This result is general enough,
and predicts tensions of the form $(e^{(c/4)\lambda + \rho},e^{(c/4)\lambda -
\rho})$, where $c$ is the (integer) constant appearing in the asymptotic
behaviour of the potential $\sim e^{c\lambda}$.

We now summarise the characteristics of the model. First, we see that the
potential behaves as $e^{4m}$.
The allowed range of parameters, given by Eq.~(\ref{m7}), is
\beq
0 < K < {1\over 2}, \,\,\, K^2 +K_0^2 ={3\over 8}.
\eeq{m15}
The IR behaviour of the metric is as in Eq.~(\ref{m4}),
and the tensions for the fundamental and the D1 strings are:
\beq
T_{F1}=e^{m+\rho}=y^{-K-K_0}; \,\,\,\, T_{D1}=e^{m-\rho}=y^{-K+K_0}.
\eeq{m16}

We are now ready to compute the Wilson loop. The computation closely parallels
that in ref. \cite{maldaloop}. The world-sheet action for a fundamental (or
D1-) string in the background~(\ref{m1}) is
\beq
S= \int d\tau d\sigma \sqrt{G_{ind}} = T\int dx
T(y)e^{\phi(y)}\sqrt{(\partial_x y)^2 + e^{2\phi(y)}}.
\eeq{m17}
The Wilson loop is obtained by minimising this action of a string whose
endpoints span a rectangle
on the boundary of the space, with one side of length $L$
in the direction $x$, and another one of length $T$ along the time axis.
 We choose the standard embedding $\sigma
=x$ and factorize the trivial integration in time. It is useful to change
coordinates in such a way that
the quark-antiquark (or
monopole-antimonopole)
 energy is
\beq
E = S/T = \int dx \sqrt{(\partial_x u)^2+f(u)}.
\eeq{m18}
The change of variable and the function $f(u)$ are given by
\beq
{\partial u\over \partial y}=T(y)e^{\phi(y)},\,\,\, f(u) = T^2(u)e^{4\phi(u)}.
\eeq{m19}
The computation of the Wilson loop for generic metrics has been widely
discussed in the literature \cite{maldaloop,rey,br1,m}. Here we
give a short review of what is expected about the Wilson loop behaviour,
once the
function $f(u)$ has been specified (more
details can be found in Appendix A). The boundary of the space is $u=+\infty$
and the asymptotic behaviour of $f(u)$ near the boundary, $f(u)\sim u^4$, is
obtained from the $AdS_5$ UV metric. The metric can be regular in the IR
region, or it can extend only up to some point $u=a$, if there is a
horizon or,
as in our case, if the metric has a singularity.
For notational simplicity we consider $a=0$.
If the function $f(u)$ has a minimum at a finite point, $\bar u$, the
fundamental string will then find energetically favourable to
end at $\bar u$ without entering in the deep IR region. Under these
circumstances,
we expect an area law for the Wilson loop \cite{w2,m}.
If instead the function $f(u)$ is monotonic in the IR, different behaviours
may be expected; they are reviewed in Appendix A. 
It may happen that
$f(u)$ behaves as $u^4$ both in the UV and in the IR, but with different
coefficients. This is the case for the RG flow  between different fixed points
considered in \cite{gppz,dz}.
Another
interesting possibility, which will be very important for us, is the behaviour
$f(u)\sim u^\gamma , 0< \gamma < 2$. In this case the string goes straight
to the point $u=0$, continues for a distance $L$ along $x$, and returns
back to $u=\infty$. Since
$f(0)=0$,
it does not cost any energy to separate the quark at $u=0$. This corresponds to
electric (or magnetic) screening~\cite{rey,br1}.

Let us apply these results to our solution. Using the coordinates in
Eq.~(\ref{m19}), we find for the function $f(u)$, in the case of a pair of
quarks and a pairs of monopoles, respectively,
\beq
f_{q\bar q}(u) \sim u^{{1-2(K+K_0)\over 5/4 -(K+K_0)}},\,\,\, f_{m\bar m}(u)
\sim u^{{1-2(K-K_0)\over 5/4 -(K-K_0)}}.
\eeq{m20}

It is easy to check that, in the allowed range of values for the constants $K$
and $K_0$, given in Eq.~(\ref{m15})\footnote{We are taking $K_0>0$. The case
$K_0<0$ is clearly related to the one discussed here by an S-duality.},
$f_{q\bar
q}$ diverges while $f_{m\bar m}\sim
u^\gamma$ with $0<\gamma <2$. This means that, in our solution,
the theory confines for all values of $K$ and $K_0$.
There is a linear potential for quarks, and
monopoles are screened.
Since confinement is obtained when the fundamental string does not probe the
far IR region, most of the  details of the solution
are therefore irrelevant. Provided that
 the function
$f(u)$ diverges in the IR, no matter how, confinement is guaranteed.

The form of the solution we found is quite similar to that found in
\cite{m}\footnote{We can easily make contact with the generic IR solution in
\cite{m}. It depends on three parameters subject to a constraint. Upon
reduction to five dimensions, the metric has, indeed, the form of
Eq.~(\ref{m4}), and does not depend on any parameter.
Two parameters are needed
to specify the tensions of the fundamental string and the Dirichlet string.
The relation between our parameters and those in \cite{m} is
$K=5{\alpha_1/2-\alpha_2\over \alpha_1-5\alpha_2}, K_0=-{\alpha_0\over
\alpha_1-5\alpha_2}$.
With this parametrisation, the quadratic constraint that we find is slightly
different from the one in \cite{m}. This is not surprising since we are dealing
with type IIB and ref. \cite{m} with type $0$. We are claiming that the form of
the solution is universal; certainly details are not.}. A difference with
ref.
\cite{m} is that we find confinement for all values of the parameters. This is
expected from the physical interpretation as a mass deformation
of N=4 super Yang-Mills.
More
general solutions, as the one in \cite{m}, for instance, which exhibit
electric confinement or screening according to different values for the
parameters,  can be easily accommodated by considering more scalars or a
different potential.

\section{Conclusions}
In this paper, we have found a very satisfying and novel form of infrared
universality for supergravity duals of strongly coupled gauge theories.
Specifically, we have found that, generically, there exists a runaway solution
of the coupled scalar-Einstein equations of motion of 5-d type IIB
gauged supergravity. Gauged 5-d supergravity is a convenient and general
framework to study deformations of N=4 super Yang-Mills theories in the
limit of large 't Hooft parameter. The runaway solution we described
is universal and independent of the detailed form of the supergravity
action. It gives a universal (singular), 5-d metric and universal formulae
for the string tensions and scalar condensates. This universality explains
why many examples of confining duals of 4-d gauge theories share common
properties. It is also encouraging in that it may mean that even for small
't Hooft parameter, some generic properties of the supergravity
approximation to gauge theories, as confinement and mass gap, can
survive string corrections.
\vskip .2in
\noindent
{\bf Acknowledgements}\vskip .1in
\noindent
M. Petrini would like to thank A. A. Tseytlin for discussions.
L. Girardello and M. Petrini are partially supported
by the European Commission TMR program ERBFMRX-CT96-0045,
wherein L. Girardello is associated to the University of Torino,
and M. Petrini to the Imperial College, London. L. Girardello is also
supported in part by INFN and MURST. M. Porrati
is supported in part by NSF grant no. PHY-9722083.

\section*{Appendix A:
Generalities About the Wilson Loop}
\renewcommand{\theequation}{A.\arabic{equation}}
\setcounter{equation}{0}
In this Appendix we review the Wilson loop calculation of
refs.~\cite{maldaloop,rey,br1,m}.

We use coordinates in which the energy of a pair of heavy quark is
\beq
E = S/T = \int dx \sqrt{(\partial_x u)^2+f(u)}.
\eeq{a1}
In the $AdS_5$ case, we have $f(u)=u^4$. The $AdS_5$
boundary is at $u=\infty$, while the singularity is at $u=0$.
The Wilson loop is obtained by minimising the corresponding action for a
string world-sheet which has a boundary at $u=\infty$. We
can take a rectangle at the boundary of the space, with one side of length $L$
in the direction $x$ and another one of length $T$ along the time axis.
We choose the standard embedding $\sigma
=x$, and we factorize the trivial integration in time.

In the conformal $AdS_5$ case, the Wilson loop is not proportional to the
boundary area since the function $f(u)$ is much smaller inside $AdS_5$ and it
is energetically favourable for the string to go inside $AdS_5$. If, for any
reason --such as a horizon, which introduces a cutoff,
as is the case of theories at finite temperature, or a
barrier-- the string cannot extend far enough from the boundary,
we expect an area law and confinement.

Consider now
a general metric corresponding to a UV non-trivial fixed point. The
boundary of the space is $u=+\infty$
and the asymptotic behaviour of the function $f(u)\sim u^4$ near the boundary
is
obtained from the $AdS_5$ UV metric. The metric may be regular in the IR
region  or extend only up to some point $u=a$ if there is a horizon or,
as in our case, if the metric has a singularity.
For notational simplicity we consider
$a=0$.

The action in Eq.~(\ref{a1}) can be minimised as in \cite{maldaloop}. Defining
$u_0$ as
the turning point of the string, we have, exactly as in \cite{maldaloop},
two equations which implicitly give $E$ as a function of $L$:
\bea
{L\over 2} &=& \int^\infty_{u_0}{du\over\sqrt{{f(u)\over
f(u_0)}(f(u)-f(u_0))}},\cr
E &=& \int^\infty_{u_0}du\left ( {\sqrt{f(u)}\over \sqrt{f(u)-f(u_0)}}-1\right
) - 1.
\eea{a2}
The derivation of these equations closely parallels
that in ref.~\cite{maldaloop}, and
it will not be repeated here. The divergent expression for the energy has been
regularized as in \cite{maldaloop} by subtracting the infinite contribution
due to the mass of the heavy quarks.

This implicit equation for $E(L)$ is hard to solve analytically. However, the
main contribution to
the energy usually comes from the region around $u_0$. The contribution from
the large-$u$ region in the integral, where $f(u)\sim u^4$, is always
subtracted, since it diverges.

Consider first the case in which $f(u)$ is monotonic. Under the assumption that
the integrals are dominated by the region around $u_0$, we get
\beq
L\sim \sqrt{{u_0\over f^{'}(u_0)}} ,\,\,\, E\sim \sqrt{{u_0f(u_0)\over
f^{'}(u_0)}}.
\eeq{a3}

It may happen that
$f(u)$ behaves as $u^4$ both in the UV and in the IR, but with different
coefficients. This is the case for the RG flow  between different fixed points
considered in \cite{gppz,dz}. In this case, for $L\rightarrow\infty$, $u_0\sim
1/L$ approaches the far IR region.
The integrals are dominated by the IR, which is still an $AdS_5$ space with a
different radius. We are back to the computation in \cite{maldaloop} with a
different radius for the space. We obviously find that, for
$L\rightarrow\infty$, $E\sim u_0\sim 1/L$, as expected in a conformal theory.

It is interesting to consider the case in which
$f(u)\sim u^\gamma , 0< \gamma < 2$. Since $L\sim u_0^{1-\gamma/2}$, we see
that for
small $u_0$ $L$ goes to zero. $L$ goes to zero also for large $u_0$, since,
due to the UV asymptotic form of $f(u)$, the integral in
Eq.~(\ref{a2}) is always convergent for $u\rightarrow\infty$. As a
consequence, $L$ is a function of $u_0$  bounded from above.
At first sight, there
is no way to probe the IR behaviour corresponding to $L\rightarrow\infty$.
However, it was argued in \cite{rey,br1} that for large enough $L$ a different
physical picture takes over. For large $L$, the string will go
to the point $u=0$ along a straight line,
proceed for a distance $L$ along $x$ and return  to $u=\infty$. Since
$f(0)=0$,
it does not cost any energy to separate the quarks at $u=0$. This corresponds
to electric (or magnetic) screening~\cite{rey,br1}.

Consider now the case that $f(u)$ is still monotonic but does not vanish in
the IR. If $f(u)\geq{\rm const}>0$,
the integral in Eq.~(\ref{a1}) is certainly
greater than $L$ and we expect an area law behaviour for the Wilson
loop \cite{w2}.
This is the case, for example, of theories at finite temperature, where
there is a horizon at a finite $u_T$.
Since $u>u_T$,
the string cannot extend far enough from the boundary.
In this case, we expect that the Wilson loop will always grow
at least as the area~\cite{w2}. This is indeed confirmed by
the expression
$f\sim {\rm const} + (u-u_T)^2$~\cite{rey,br1}. We see that $u_0$ will
approach the horizon.

Beside the existence of a horizon, there is a second mechanism that prevents
 the
 string from going far enough from the boundary. If
the function
$f(u)$ diverges near $u=0$, it will create a barrier,
giving rise to a confining behaviour for the Wilson loop.
$f(u)$
will have a minimum at a finite point $\bar u$. Assuming $f(\bar u)>0$, we
have that $f(u)\geq f(\bar u)>0$. The
fundamental string will then find energetically favourable to
end at $\bar u$ without entering in the far IR region. Under these
circumstances,
we expect an area law for the Wilson loop \cite{w2,m}.
We can explicitly check the behaviour of the Wilson loop. $L$ automatically
diverges if  $u_0$ approaches $\bar u$. Suppose that $f^{'}(\bar u)=0,
f^{''}(\bar u)\ne 0$ --the case in which the first non-zero derivative is the
$k$-th one is completely analogous. From Eqs.~(\ref{a2}) we have:
\beq
{L\over 2}\sim {\rm const}\int^\infty_{u_0}{du\over u-\bar
u}\rightarrow\infty,\,\,\, E\sim {\rm const}\int^\infty_{u_0}{du\over
u-\bar u} \sim L
\eeq{a4}
and we see that, generically, $E\sim L$.

\section*{Appendix B: Conventions and Useful Formulae for the Lagrangian}
\renewcommand{\theequation}{B.\arabic{equation}}
\setcounter{equation}{0}
In this Appendix we give the relevant formulae needed to compute
 the Lagrangian in Eq. (\ref{m9}). Since the tools for computing the
potential have been already described in details in
\cite{grw2} and recently reviewed in \cite{dz,kpw},
we do not aim to be self-contained but just to sketch the outline of the
computation. The  reader may refer to the papers mentioned above for
more details. Ref.~\cite{kpw}, in particular, gives a nice short review of
the philosophy behind these computations.

The five-dimensional Lagrangian for the scalars is written in terms of
the $27 \times 27$ matrix $U$ parametrising the coset $E_6/USp(8)$
\beq
L = \sqrt{-g}\left[{R\over 4} - {1\over 24}\Tr (U^{-1}\partial U)^2 + V(U)
\right].
\eeq{aa1}
In a unitary gauge, $U$ can be written as
\beq
U=\exp{X},
\eeq{aa2}
where $X= \sum_{a} \lambda_a T_a$ is given by the $42$ generators of $E_6$
that do not belong to $USp(8)$: these $42$ independent parameters
correspond to the $42$ scalars.
The precise form of the generator corresponding to a given
scalar can be worked out using the global and local symmetries of the problem,
and the fact that $U$ maps an element of the
representation $\underline{27}$ of $E_6$ into itself.
More precisely, one has to remember that only the group
$SU(4)\times SL(2;R)$ is a symmetry of the Lagrangian. The $42$ scalars
then decompose according to
\beq
42 \rightarrow 20'_{(0)} + 10_{(-2)} + \bar{10}_{(2)} + 1_{(4)} + 1_{(-4)},
\eeq{aa3}
while the vectors in the $\underline{27}$ decompose as
\beq
27 \rightarrow 15_{(0)} + 6_{(2)} + 6_{(-2)}.
\eeq{aa4}
The subscripts denote the charges of the $U(1)$ factor in $SL(2;R)$.

The general parametrisation for $U$ is given by Eqs.~(A.36) of
\cite{grw2}, with the following conventions for the indices:
the $\underline{27}$ of $E_6$ is represented by a couple of antisymmetric
symplectic-traceless indices $A,B$,
running form $1$ to $8$, and, in the  $SU(4)\times
SL(2;R)$ basis, it decomposes as
\bea
27 & \rightarrow & 15 + (6,2),\cr
[AB] & \rightarrow & ([IJ],\, I\alpha)
\eea{aa5}
with $I,J=1,\dots,6$ indices of $SU(4)$ and $\alpha=1,2$ indices of $SL(2;R)$.
The brackets mean that the indices are antisymmetrized.

If one is interested, as we are, in breaking the $SU(4)$ factor to
$SU(3)\times U(1)$, a more suitable parametrisation for $U$ is given in
\cite{dz}, with the following basis for the $\underline{27}$:
\beq
27 \rightarrow \left(1_{(0,0)}, 3_{(4,0)}, \bar{3}_{(-4,0)}, 8_{(0,0)},
3_{(-2,2)},
\bar{3}_{(2,2)}, 3_{(-2,-2)}, \bar{3}_{(2,-2)}\right).
\eeq{aa6}
This time the subscripts indicate the charges under the
$U(1)\subset SU(4)$ and the $U(1)\subset SL(2;R)$, respectively.

In our example, we turn on the dilaton field, $\rho$, and the scalars $m$
in the $\underline{6}_{(2,-2)}$ of $SU(4)$. From ref.~\cite{dz},
the matrix $U$ has the form
\beq
U=e^{X_0} e^{X},
\eeq{aa7}
where
\beq
X_0=\left( \begin{array}{cccccccc}
          0 & 0 & 0 & 0 & 0    & 0   & 0    &  0 \\
          0 & 0 & 0 & 0 & 0    & 0   & 0    &  0 \\
          0 & 0 & 0 & 0 & 0    & 0   & 0    &  0 \\
          0 & 0 & 0 & 0 & 0    & 0   & 0    &  0 \\
          0 & 0 & 0 & 0 & 0    & 0   & \rho e^{i\alpha} & 0 \\
          0 & 0 & 0 & 0 & 0    & 0   & 0    & \rho e^{i\alpha}\\
          0 & 0 & 0 & 0 & \rho e^{-i\alpha} & 0   & 0    & 0 \\
          0 & 0 & 0 & 0 & 0    &\rho e^{-i\alpha} & 0    & 0
          \end{array} \right), \qquad
X=\left( \begin{array}{cccccccc}
          0 & 0     & 0 & 0 &   0    & 0   & 0       &  0 \\
          0 & 0     & 0 & 0 &   0    & m   & 0       &  0 \\
          0 & 0     & 0 & 0 &   0    & 0   & \bar{m} &  0 \\
          0 & 0     & 0 & 0 &   0    & 0   & 0       &  0 \\
          0 & 0     & 0 & 0 &   0    & 0   & 0       & 0 \\
          0 & \bar{m}     & 0 & 0 &   0    & 0   & 0       & 0 \\
          0 & 0     & m & 0 &   0    & 0   & 0       & 0 \\
          0 & 0     & 0 & 0 &   0    & 0   & 0       & 0
          \end{array} \right).
\eeq{aa8}

The potential and the tensions of the fundamental string and of the
D1-string are all given in terms of the vielbein $V_{AB}\null^{ab}$,
where $a,b$ are a couple of antisymmetric symplectic-traceless indices
$a,b=1,...,8$, and representing
the $\underline{27}$ of $USp(8)$.
This field, being an element of $E_6/USp(8)$, carries both the indices
$A,B=1,\dots,8$ of the $\underline{27}$ of $E_6$ and the
indices $a,b=1,\dots,8$ of the $\underline{27}$ of $USp(8)$.
One can pass from the $E_6$ basis to the $USp(8)$ one using the $SO(7)$
gamma matrices, $\Gamma$, defined in the Appendix of \cite{grw2}.


In our basis the vielbein is simply obtained by multiplying on the right the
matrix
$U$  by the following vector of gamma matrices:
\beq
\left(\frac{\gamma_{i\bar{\jmath}}}{4\sqrt{2}},
\frac{\epsilon_{ijk}\gamma_{\bar{\jmath}\bar{k}}}{4\sqrt{2}},
\frac{\epsilon_{\bar{\imath}\bar{\jmath}\bar{k}}\gamma_{jk}}{4\sqrt{2}},
\frac{\gamma_{i\bar{\jmath}}}{4\sqrt{2}},
\frac{\gamma_{i}(1-\Gamma_0)}{4},
\frac{\gamma_{\bar{\imath}}(1-\Gamma_0)}{4},
\frac{\gamma_{i}(1+\Gamma_0)}{4},
\frac{\gamma_{\bar{\imath}}(1+\Gamma_0)}{4}\right).
\eeq{aa11}

Here the complex indices $i,\bar{\imath}$ run from $1$ to $3$, and the new
complex gamma matrices, $\gamma_{i}$ and $\gamma_{\bar{\imath}}$, are related
to
the real matrices of \cite{grw2} by
\bea
\gamma_{1}&=&\frac{\Gamma_1+i\Gamma_2}{\sqrt{2}}, \dots\\
\gamma_{\bar{1}}&=&\frac{\Gamma_1-i\Gamma_2}{\sqrt{2}}, \dots.
\eea{aa12}
The computation of the potential is now only a matter of straightforward
gamma matrix algebra: just plug $V_{AB}\null^{ab}$ in the expression for the
potential \cite{grw2}
\beq
V = -\frac{1}{32}g^2 \left[2W_{ab}W^{ab} - W_{abcd}W^{abcd}\right],
\eeq{aa13}
where
\beq
W_{abcd}=\epsilon^{\alpha \beta}\eta^{IJ}V_{I\alpha ab}V_{J\beta ab}.
\eeq{aa14}
In these equations the $USp(8)$ indices are raised and lowered with the
matrix $\Gamma_0$, as shown in ref.~\cite{grw2}.

Two features of the potential are worth mentioning. First, the
potential does not depend on the dilaton, since the factors involving
$\exp{X_0}$ cancel out. Moreover, the potential is quartic in the
matrix $U$; therefore, one expects an asymptotic dependence on the field $m$
of the form
\beq
V\sim e^{4m}.
\eeq{aaa14}

Similarly, the kinetic terms for the $12$ antisymmetric tensors
$B_{\mu\nu}^{I\alpha}$ are \cite{grw2}
\beq
-\frac{1}{8}B_{\mu\nu ab}B^{\mu\nu ab} +
\frac{1}{8g}\epsilon^{\mu\nu\rho\sigma\tau}
\eta_{IJ}\epsilon_{\alpha\beta}
B_{\mu\nu}^{I\alpha}D_{\rho}B_{\sigma\tau}^{J\beta},
\eeq{aa15}
with $B_{\mu\nu}^{ab}=B_{\mu\nu}^{I\alpha}V_{I\alpha}^{ab}$.

Consider here the simpler case where only the dilaton is turned
on. This has some interest in itself since
for the theory at
the N=8 supersymmetric vacuum all scalars but the dilaton are zero.
The matrix $U$ reduces to
\beq
U=e^{X_0}
\eeq{aa16}
with $X_0$ given in Eq. (\ref{aa8}), and we obtain for $V_{AB}\null^{ab}$
\bea
V_{I+}\null^{ab}&=&\frac{1}{4}\left(\cosh\rho + \sinh\rho e^{i\alpha}\right)
\Gamma_I
-\frac{1}{4}\left(\cosh\rho - \sinh\rho e^{i\alpha}\right)\Gamma_I\Gamma_0,\\
V_{I-}\null^{ab}&=&\frac{1}{4}\left(\cosh\rho +
\sinh\rho e^{-i\alpha}\right)\Gamma_I
+\frac{1}{4}\left(\cosh\rho - \sinh\rho e^{-i\alpha}\right)\Gamma_I\Gamma_0.
\eea{aa17}

In the  real
basis for $SL(2;R)$ we find, considering for simplicity a zero axion field
$\alpha$,
\bea
V_{I1}\null^{ab}&=&\frac{1}{2\sqrt{2}}
\left(\cosh\rho + \sinh\rho\right)\Gamma_I,
\cr
V_{I2}\null^{ab}&=&
\frac{i}{2\sqrt{2}}\left(\cosh\rho - \sinh\rho\right)\Gamma_I\Gamma_0.
\eea{aa18}
We see that the matrix $A_{I\alpha,J\beta}=V_{I\alpha}\null^{ab}V_{J\beta ab}$,
which appears in the Lagrangian in the quadratic terms for
$B_{\mu\nu}^{I\alpha}$,
is diagonal both in the $SO(6)$ and the $SL(2;R)$ indices. 
The square 
of the tensions can be read from the restricted matrices $A_{I1,J1}$ and
$A_{I2,J2}$. 
 Not surprising, the eigenvalues 
are
$e^{2\rho},e^{-2\rho}$. 

It is quite tedious, but straightforward to compute the matrix $A$ for 
non-zero $m$. The result is (for zero axion field),
\beq
A_{I\alpha,J\beta}= {\cosh (2m) +1\over 2}\Delta_{\alpha\beta}\delta_{IJ}
+ {\cosh (2m) - 1\over 2}\epsilon_{\alpha\beta}J_{IJ},
\eeq{aaaauffa}
where $J_{IJ}={\rm diag}(i\sigma_2, i\sigma_2, i\sigma_2)$ is a $6\times 6$ 
matrix, and $\Delta={\rm diag\,}[\exp(2\rho),\exp(-2\rho)]$. 
The off-diagonal terms are irrelevant for our computation, since they
modify the mass matrix of the antisymmetric tensors, 
but not their kinetic term. 
The result for the tensions, is
$e^{m+\rho},e^{m-\rho}$. We see that the contribution of the scalar is the
same for both type of tensions, and, due to the fact that $A$ is quadratic in
$U$, has the asymptotic form $e^{m}$. This result applies to other scalars
as well.


\begin{thebibliography}{6666666666}
\bibitem{'t} G. 't Hooft, Nucl. Phys. B72 (1974) 461
\bibitem{malda} J. Maldacena, Adv. Theor.
Math. Phys. 2 (1998) 231, hep-th/9711200.
\bibitem{gkp} S.S. Gubser, I.R. Klebanov and A.M. Polyakov, Phys.Lett. B428
(1998) 105, hep-th/9802109.
\bibitem{w1} E. Witten, Adv. Theor. Math. Phys. 2 (1998) 253,
hep-th/9802150.
\bibitem{imsy} N. Itzhaki, J. Maldacena, J. Sonnenschein and S. Yankielowicz,
Phys. Rev. D58 (1998) 046004, hep-th/9802042.
\bibitem{w2} E.Witten, Adv. Theor. Math. Phys. 2 (1998) 505, hep-th/9803131.
\bibitem{r} J. Russo, hep-th/9808117; C. Csaki, Y. Oz and J.Russo,
Phys. Rev. D59 (1999) 065008, hep-th/9810186.
\bibitem{ks} A. Kehagias and K. Sfetsos, hep-th/9902125; S. Gubser,
hep-th/9902155.
\bibitem{kt} I.R. Klebanov and A.A. Tseytlin, hep-th/9811035; hep-th/9812089;
hep-th/9901101.
\bibitem{m} J.A. Minahan, hep-th/9902074.
\bibitem{poly} A. Polyakov, Nucl. Phys. Proc. Suppl. 68 (1998) 1, 
hep-th/9711002.
\bibitem{poly2}  A. Polyakov, hep-th/9809057.
\bibitem{krv} H. J. Kim, L. J. Romans and P. van Nieuwenhuizen, Nucl. Phys.
B242 (1984) 377.
\bibitem{grw1} M. G\"unaydin, L.J. Romans and N.P. Warner, Phys. Lett. B154
(1985) 268; M. Pernici, K. Pilch and P. van Nieuwenhuizen, Nucl. Phys. B259
(1985) 460.
\bibitem{grw2} M. G\"unaydin, L.J. Romans and N.P. Warner, Nucl. Phys. B272
(1986) 598.
\bibitem{bf} P. Breitenlohner and D.Z. Freedman, Ann. Phys. 144 (1982) 249.
\bibitem{gppz} L. Girardello, M. Petrini, M. Porrati and A. Zaffaroni,
JHEP 12 (1998) 022, hep-th/9810126.
\bibitem{dz} J.Distler and F.Zamora, hep-th/9810206.
\bibitem{kpw} A. Khavaev, K. Pilch and N.P. Warner, hep-th/9812035.
\bibitem{df} E. D'Hoker and D.Z. Freedman, hep-th/9811257.
\bibitem{dn} B. de Wit and H. Nicolai, Nucl. Phys. B281 (1987) 211; B. de Wit,
H. Nicolai and N.P. Warner, Nucl. Phys. B255 (1984) 29.
\bibitem{ht} S.W. Hawking and N. Turok, Phys. Lett. B425 (1998) 25,
hep-th/9802030.
\bibitem{maldaloop} J. M. Maldacena, Phys. Rev. Lett. 80 (1998) 4859,
hep-th/9803002.
\bibitem{mina} J.A. Minahan, hep-th/9811156.
\bibitem{ll} L.D. Landau and E.M. Lifshitz, 
{\em Quantum Mechanics} (Pergamon Press,
Oxford, III edition, 1976).
\bibitem{flz} S. Ferrara, M.A. Lled\'o and 
A. Zaffaroni, Phys. Rev. D58 (1998) 105029, hep-th/9805082.
\bibitem{rey} S. J. Rey and  J. Yee, hep-th/9803001;  S. J. Rey, S. Theisen and
J. Yee, Nucl. Phys. B527 (1998) 171, hep-th/9803135.
\bibitem{br1} A. Brandhuber, N. Itzhaki, J. Sonnenschein and S. Yankielowicz,
Phys.Lett. B434 (1998) 36, hep-th/9803137; JHEP 9806 (1998) 001,
hep-th/9803263.
\end{thebibliography}
\end{document}